\documentclass[dvips]{acta}
\usepackage{supertabular,lscape,epsfig}
\usepackage{amssymb}
\usepackage{amsmath}
\usepackage{graphicx}

\SetPages{0}{0}

\SetVol{58}{2008}

\begin{document}

\begin{Titlepage}
\Title{P13 an ULX that is a potential progenitor of merging BH-NS system}
\Author{Beldycki, B$^{3}$, Belczynski, K$^{1,2}$}{$^{1}$ Astronomical 
Observatory, Warsaw University, al. Ujazdowskie 4, 00-478 Warsaw, Poland

$^{2}$ Warsaw Virgo Group

$^{3}$ Nicolaus Copernicus Astronomical Center, Polish Academy of Sciences, 
Bartycka 18, 00-716 Warsaw, Poland
}

\Received{Month Day, Year}
\end{Titlepage}

\Abstract{We have studied the future evolution of a recently discovered ULX source P13 
in NGC7793. This source was shown to contain a 5-15$\MS$ black hole and a massive
18-23$\MS$ B9Ia companion on a 64 day orbit. For low black hole mass (5-10$\MS$) and
high companion mass $\gtrsim 20 \MS$ the binary is predicted to initiate a common 
envelope evolution in near future and significantly decrease the orbital separation 
(6 hr orbit). This leads to a high probability ($\sim70\%$) of the system surviving a 
supernova explosion that will form a neutron star out of the companion.
About one third of the surviving BH-NS systems will merge within Hubble time and be a 
source of high frequency gravitational radiation. We estimate that the chances of 
detection of BH-NS systems with advanced LIGO/Virgo that form via P13-evolutionary 
channel are at the level of 0.1 yr$^{-1}$ with wide range of allowed probability 
$0 \div 0.6$yr$^{-1}$. This is the fourth empirical estimate of BH-NS merger 
rate.}
{binaries: close -- stars: evolution, neutron stars -- gravitation}

\section{INTRODUCTION}
Universe hosts black holes (BHs) and neutron stars (NSs) in various binary configurations.
Many of these binaries show significant X-ray activity that suggests mass transfer and 
accretion onto a compact object. Recently observed P13 system is an extra-galactic X-ray 
source. This X-ray source is unique because it is the first ULX with the reliable mass 
estimate of compact object: apparently a stellar-origin black hole (Motch et al. 2014). 
Pietrzynski et al. (2010) have determined distance to NGC7793 (P13 host galaxy) using 
Cepheids from variable stars discovered in Wide-Field Imaging Survey. P13 is a high mass 
X-ray binary (HMXB) hosting a  stellar-origin black hole (5-15$\MS$) in 64 day orbit 
around a massive (18-23$\MS$) B9Ia star. The companion is in the mass range for neutron 
star formation (Belczynski et al. 2008). We select this system to investigate a population 
of potential Gravitational Wave (GW) sources: Black Holes + Neutron Star (BH-NS) mergers. 
Over the past few years three studies have been published in which authors examined value 
of emipiral merger rate for BH-NS systems. In 2011, Belczynski et al. (2011) obtained the 
BH-NS merger rate of $\sim$0.001 yr$^{-1}$ for Cyg X-1 system. In 2013, Belczynski et al.
(2013) obtained the BH-NS empirical merger rate of$\sim$0.1 yr$^{-1}$ for Cyg X-3 system. 
And in 2015, Grudzinska et al. (2015) calculated empirical BH-NS merger rate of 
$\sim$0.2 yr$^{-1}$ for MWC 656 system. These systems host BHs with masses in range 
$5-15\MS$ and companion stars with mass range from $7\MS$ to $20\MS$. P13 is therefore 
rather similar.

\section{ESTIMATES}

\subsection{The evolution of P13}

To investigate the future evolution of P13 system we employ the binary population 
synthesis code, $\textbf{StarTrack}$ (Belczynski et al. 2008a). We use average values  
of estimated orbital parameters: $P_{\rm orb} = 64d$ and eccentricity $e=0.34$. 
We investigate three cases of binary configurations (dependent on component masses) which 
undergo different evolutionary paths. In case 1, binary system experiences only common 
envelope phase (CE) while donor is a Hertzsprung gap star. In case 2, binary system 
experiences only Roche Lobe Overflow (RLOF) while donor is a Hertzsprung gap star. In 
case 3, binary system experiences both CE and RLOF phase; CE is encountered while
donor is a Hertzsprung gap star and RLOF occurs while donor is a naked evolved Helium 
star. 

{\bf Case 1.}
We start evolutionary calculation, while the more massive star in binary has already 
formed a BH with mass $M_{\rm BH} = 10\MS$. This takes about 5 Myr. Therefore, we 
assume that the massive companion in P13 is a Main Sequence (MS) star that has already 
undergone significant nuclear burning and its mass is $M_{\rm comp} = 25\MS$.  
Following observations we assume orbit with $P_{\rm orb} = 64d$ ($a=220\RS$) and 
eccentricity $e=0.34$.

At $t=7.4$ Myr the companion evolves off MS. At this point companion star lost 
$\sim0.7\MS$ and orbit became wider ($P_{\rm orb} = 66d$, $a=224\RS$), but tidal
forces were not able to reduce eccentricity ($e=0.34$).

During Hertzsprung Gap (HG) phase the system undergoes CE phase. Here we treat CE phase 
with standard energy balance (Webbink 1984). We assume the efficient conversion of orbital 
energy into the envelope ejection ($\alpha_{\rm CE} = 1$). The binding energy of star 
envelope is obtained with physical stellar models (binding energy parameter $\lambda = 
0.05$; Dominik et al. 2012). During the CE phase part of envelope is accreted onto BH 
(it increases its mass from $10\MS$ to $10.15\MS$) and the rest is ejected from the binary 
system. Companion star loses a significant fraction of its mass $M_{\rm comp} = 7.2\MS$
after CE and becomes a massive helium core or Wolf-Rayet (WR) star. Orbital period
decays to just $P_{\rm orb} = 0.24d$ ($a=4.2\RS$) and becomes circular. 

After CE, from $t=7.4$ Myr to $t=8.38$ Myr the WR star wind is accreted onto BH and 
the system enters HMXB phase. The WR star mass decreases to $6.15\MS$ and
in response $P_{\rm orb} = 0.32d$ ($a=4.5\RS$).

Evolution ends at $t=8.38$ Myr with supernova explosion (SN) and NS formation: 
$M_{\rm NS} = 1.56\MS$. We adopt natal kicks with uniform distribution of orientations 
and with a 1D Maxwellian velocity distribution with $\sigma$ = 265 km s$^{-1}$. This 
distribution was measured for Galactic single pulsars (Hobbs et al. 2005). Due to the 
fact that binary is on close orbit during the SN ($P_{\rm orb} = 0.32d$, $a=4.5\RS$) 
the system has significant chance ($\sim 38\%$) to form close binary system which will 
merge within Hubble time. During evolution the BH has accreted $\Delta M = 0.15\MS$. 
If the birth BH spin was $a=0.0$, $0.5$, $0.9$ then it was increased to $a = 0.055$, 
$0.531$ and $0.909$, respectively.

{\bf Case 2.} 
We start evolutionary calculation, while the more massive star in binary has already 
formed a BH with mass $M_{\rm BH} = 15\MS$. This takes about 4.49 Myr. Therefore, we 
assume that the massive companion in P13 is a MS star that has already 
undergone significant nuclear burning and its mass is $M_{\rm comp} = 22.5\MS$.  
Following observations we assume orbit with $P_{\rm orb} = 64d$ ($a=225\RS$) and 
eccentricity $e=0.34$.

At $t=8.28$ Myr companion star evolves off MS. At this point companion 
star lost $\sim0.5\MS$ and orbit became wider ($P_{\rm orb} = 65d$, $a=228\RS$), but 
tidal forces were not able to reduce eccentricity $e=0.34$.

During HG phase companion star expands and fills its Roche Lobe ($R_{\rm 2} = R_{\rm 2, lobe} = 63\RS$) 
and the RLOF proceeds. This time due to less extreme mass ratio (as contrasted with 
case 1) the mass transfer is stable (no CE evolution). RLOF starts at $t=8.29$ Myr 
and lasts about $0.005$ Myr. We assume non-conservative evolution throughout the 
RLOF with accretion onto BH Eddington limited. We assume that the lost material takes 
away angular momentum specific to the accretor. During the RLOF phase part of mass is 
acreted onto BH (it increases its mass from $15\MS$ to $17.75\MS$). Companion star mass 
decreases from $22\MS$ to $6.44\MS$ and it becomes a massive helium core or WR star. 
Orbit period expands to $P_{\rm orb} = 213d$ ($a=434\RS$) and becomes circular.

The WR star mass decreases to $5.6\MS$ ($P_{\rm orb} = 328d$, $a=571\RS$) due to 
intensive wind mass loss. Evolution ends at $t=9.32$ Myr with SN and NS formation: 
$M_{\rm NS} = 1.46\MS$. Due to the fact that binary is on wide orbit during the SN 
($P_{\rm orb} = 328d$, $a=571\RS$) the system has only a small chance ($\sim 0.01\%$) 
to form close binary system which will merge within Hubble time. During evolution 
the BH has accreted a total of $\Delta M = 2.75\MS$. If the birth BH spin was $a=0.0$, 
$0.5$, $0.9$ then it was increased to $a = 0.487$, $0.771$ and $0.97$, respectively. 

{\bf Case 3.}
We start evolutionary calculation, while the more massive star in binary has already 
formed a BH with mass $M_{\rm BH} = 10\MS$. This takes about 5 Myr. Therefore, we 
assume that the massive companion in P13 is a MS star that has already 
undergone significant nuclear burning and its mass is $M_{\rm comp} = 22.5\MS$.  
We assume orbit with $P_{\rm orb} = 64d$ ($a=214\RS$) and $e=0.34$.

At $t=8.28$ Myr the companion evolves off MS. At this point companion star lost 
$\sim0.5\MS$ and orbit became wider ($P_{\rm orb} = 66d$, $a=217\RS$) with 
$e=0.34$.

During HG phase the system undergoes CE phase. During the CE phase part of envelope is 
accreted onto BH (it increases its mass from $10\MS$ to $10.17\MS$) and the rest is 
ejected from the binary system. Companion star loses a significant fraction of its 
mass $M_{\rm comp} = 6.31\MS$ after CE and becomes a massive WR star. Orbital period
reduces to just $P_{\rm orb} = 0.24d$ ($a=4.2\RS$) and the orbit becomes circular.

Next, the WR star mass decreases to $5.37\MS$ due to wind mass loss and in response 
orbit changes to $P_{\rm orb} = 0.27d$ ($a=4.4\RS$). At $t=9.32$ Myr WR star expands 
slightly and fills its Roche Lobe ($R_{\rm 2} = R_{\rm 2, lobe} = 1.45\RS$). 
The stable RLOF lasts about $0.001$ Myr. During the RLOF phase part of mass is acreted 
onto BH (it increases its mass from $10.17\MS$ to $10.47\MS$). WR star mass decreases 
from $5.37\MS$ to $3.93\MS$. Orbit period expands to $P_{\rm orb} = 0.61d$ ($a=6.95\RS$).

Evolution ends at $t=9.33$ Myr (just after RLOF) with SN and NS formation: $M_{\rm NS} = 1.45\MS$. 
Due to the fact that binary is on close orbit during the SN ($P_{\rm orb} = 0.61d$, 
$a=6.95\RS$) the system has significant chance ($\sim 22\%$) to form close binary system 
which will merge within Hubble time. During evolution the BH has accreted $\Delta 
M = 0.47\MS$. If the birth BH spin was $a=0.0$, $0.5$, $0.9$ then it was increased 
to $a = 0.15$, $0.59$ and $0.92$, respectively.

\subsection{Rate Estimates}

P13 has been observed because of its unusually strong X-ray emission. Such systems are X-ray 
active as long as evolution of the secondary component has not come to the end and it can feed 
mass to a compact object. Entire evolution of the secondary object lasts for about 9 Myr 
(e.g., case 3). We employ this as the upper limit, i.e. conservative estimate for the BH-NS 
detection rate. More reliable estimate for X-ray phase is only 4 Myr (since it takes about 
5 Myr to form BH). Such reduced estimate would have resulted in higher (by factor of $\sim 2$) 
detection rate.

Distance within which LIGO/VIRGO can detect GW from a given source (e.g. BH-NS merger) is 
given by
\begin{equation}
\label{1}
d_{\rm dco} = d_{\rm 0} \bigg(\frac{M_{\rm c,dco}}{M_{\rm c,nsns}}\bigg)^{5/6},
\end{equation}
where $M_{c,dco} = (M_{\rm BH}M_{\rm NS})^{3/5}(M_{\rm BH} + M_{\rm NS})^{-1/5}$ is a chirp mass 
of the BH-NS system, $M_{\rm c,nsns} = M_{\rm 1}^{3/5}M_{\rm 2}^{3/5}(M_{\rm 1}+M_{\rm
2})^{-1/5}=1.2\MS$ is a chirp mass for double neutron star system with 
$M_{\rm 1}=M_{\rm 2}=1.4\MS$, $d_{\rm 0}$ is a horizon distance for NS-NS merger and 
for advanced LIGO/VIRGO sensitivity it is expected to be $d_{\rm 0} = 450$ Mpc. For example, 
in case 3, we have $M_{\rm c,dco}=3.2\MS$ that gives $d_{\rm dco} = 1.0$ Gpc. 

We observe only one P13 system. It is located at distance 3.4 Mpc. If we assume that similar 
systems to P13 are uniformly distributed in space and form with equal probability in time we 
can estimate how many of them are merging within range of Advanced LIGO/VIRGO detectors. 
The LIGO/VIRGO detection rate can be estimated from: 
\begin{equation}
\label{2}
R_{\rm LIGO} =  \frac{f_{\rm close}}{t_{\rm life}}\bigg(\frac{(d_{\rm
c,dco}/f_{\rm pos})}{d_{\rm P13}}\bigg)^{3},
\end{equation}
where $f_{\rm close}$ is a probability that system survives SN explosion and form close binary system
 which merge within Hubble Time, $t_{\rm life}$ is an evolution time for P13, $d_{\rm c,dco}$ is a 
distance within which LIGO/VIRGO can detect GW from a given source, $f_{\rm pos} = 2.26$ is a 
correction factor that takes into account the non-uniform pattern of detection and random 
sky location and orientation of sources (Finn, 1996), and $d_{\rm P13}$ is a distance to
P13 system. For case 3, we obtain th value of empirical detection rate at level 
$R_{\rm LIGO} =  0.123$ yr$^{-1}$. Other cases (black hole mass $M_{\rm BH} = 5,10,15 M_{\MS}$, 
companion star mass $M_{\rm comp} = 20,22.5,25\MS$) are shown in Table 1.

\section{DISCUSSION}

We obtain detection rate at level $10^{-4}\div 10^{-5}$ yr$^{-1}$ in case 2 (case without 
CE phase). In case 1 and 3 (progenitor systems undergo CE phase) detection rates are
much higher $0.1\div0.6$ yr$^{-1}$. The highest value of detection rate was obtained for 
black hole mass $5\MS$ and companion star mass $22.5\MS$. In previous papers authors 
obtained value of emipirical merger rate for BH-NS progenitors at level $R_{\rm LIGO} = 
2.8 \times 10^{-2} \div 0.4$ yr$^{-1}$ for Cyg X-1 system (Belczynski et al. 2011), 
$R_{\rm LIGO} = 0.09 \div 0.15$ yr$^{-1}$ for Cyg-X3 system (Belczynski et al. 2013), 
and $R_{\rm LIGO} = 0.0 \div 0.187$ yr$^{-1}$ for MWC 656 system (Grudzinska et al. 2015).
Our results are consistent with the previous estimates.

For our cases 1 and 3, where we find small, but significant, detection rates
the increase of BH spin is negligible due to low mass accretion in CE.
Therefore, if BH-NS are detected, and if they were formed in P13-channel the
BH spins are birth spins. 
In case 2 we expect significant accretion onto BH during stable RLOF, but
the detection rates are negligibly small. 
This is similar to the evolution of BH-BH binaries, for which BH spins are
expected to be very close to birth spins as well (Belczynski et al. 2015). 

The nature of compact object in P13 is not established. It was {\em assumed}
by Motch et al. (2014) that the compact object is more massive than $5\MS$ and
therefore it must be a BH. However, it cannot be excluded that the mass is
lower. If the compact object in P13 is a NS and not a BH then evolution follows 
quite a different path. 
If we replace BH with NS in P13 system, then it will be imposible to get a close 
binary system with two NSs which will merge within Hubble Time. Close NS-NS
formation is not expected as the supernova mass loss will be greater than half of 
the P13 total binary mass before SNa. This would cause the binary disruption
even without any natal kick. Such evolution {\em always} results in a null 
detection rate.

Evolutionary channels that generate high detection rates ($>0.1$ yr$^{-1}$)
involve CE phase. However, it is not at all clear whether CE can be
initiated (Pavlovskii and Ivanova 2014/2015) or survived (Belczynski et al. 2007) 
if a donor star is a Hertzsprung gap star: as in {\em all} our evolutionary
scenarios that result in high detection rates (cases 1 and 3). If this is taken 
into account then our major BH-NS formation sequences from P13-like progenitors 
will not produce any BH-NS mergers. 

To summarize, we have considered consequences of the future evolution of ULX P13. 
We have shown that it is potentially possible that such binary may form
BH-NS merger and that associated detection rate with advanced LIGO/Virgo
is small, but significant. 
However, we cannot exclude the possibility, that this type of binary will not 
form a double compact object merger. Therefore, our predicted detection
rate of BH-NS mergers varies from no detections to $0.6$ yr$^{-1}$. This is the 
fourth empirical estimate of BH-NS merger detection rate.



\newpage
\begin{center}
\MakeTable{c c c c c c c c}{12.5cm}{The Fate of P13}
{\hline
 \hline
$M_{\rm BH} + M_{\rm comp}$ & Outcome$^{\textrm{a}}$ & $f_{\rm close}$ $^{\textrm{b}}$ 
& $M_{\rm c,dco}$$^{\textrm{c}}$ & $t_{\rm life}$$^{\textrm{d}}$ & $d_{\rm c,dco}$$^{\textrm{e}}$ 
& $R_{\rm LIGO}$$^{\textrm{f}}$ \\
\hline
\\
5 + 20.0 $M_{\odot}$ & (5.4 + 1.28) & 0.33 & 2.1 & 2.0 & 327 & 0.14\\
5 + 22.5 $M_{\odot}$ & (5.44 + 1.41) & 0.39 & 2.3 & 0.7 & 343 & 0.57\\
5 + 25.0 $M_{\odot}$ & (5.45 + 1.55) & 0.40 & 2.4 & $< 0.7$ & 359 & $< 0.57$\\
\\
10 + 20.0 $M_{\odot}$ & (12.3 + 1.34) & $2.6 \times 10^{-4}$ & 3.1 & 5.6 & 449 
& $1.07 \times 10^{-4}$ \\
10 + 22.5 $M_{\odot}$ & (10.45 + 1.56) & 0.22 & 3.2 & 4.3 & 456 & 0.12 \\
10 + 25.0 $M_{\odot}$ & (10.15 + 1.45) & 0.38 & 3.0 & 3.3 & 436 & 0.23 \\
\\
15 + 20.0 $M_{\odot}$ & (17.35 + 1.39) & $10^{-4}$ & 3.7 & 6.0 & 515 & $7.6\times
 10^{-5}$\\
15 + 22.5 $M_{\odot}$ & (17.67 + 1.46) & $1.3 \times 10^{-4}$ & 3.8 & 4.8 & 531 
& $1.0\times 10^{-4}$ \\
15 + 25.0 $M_{\odot}$ & (17.82 + 1.57) & $2.08 \times 10^{-4}$ & 4.0 & 3.8 & 552
 & $2.2\times 10^{-4}$ \\
\hline
\hline
\multicolumn{7}{p{12cm}}{$^{a}$ Close BH-NS binary system with the mass of each 
object in solar mass units. $^{b}$ Propability that system survives SN explosion 
and form close binary system which marge within Hubble Time. $^{c}$ Chirp mass 
of our system in solar mass units. $^{d}$ Time evolution of P13 system in Myr. 
$^{e}$ Maximum detection range for advanced LIGO/VIRGO detectors in Mpc. $^{f}$ 
Estimated merger rate per year.}
}

\end{center}

\newpage

\begin{figure}[htb]
\includegraphics[width=\textwidth,height=\textwidth]{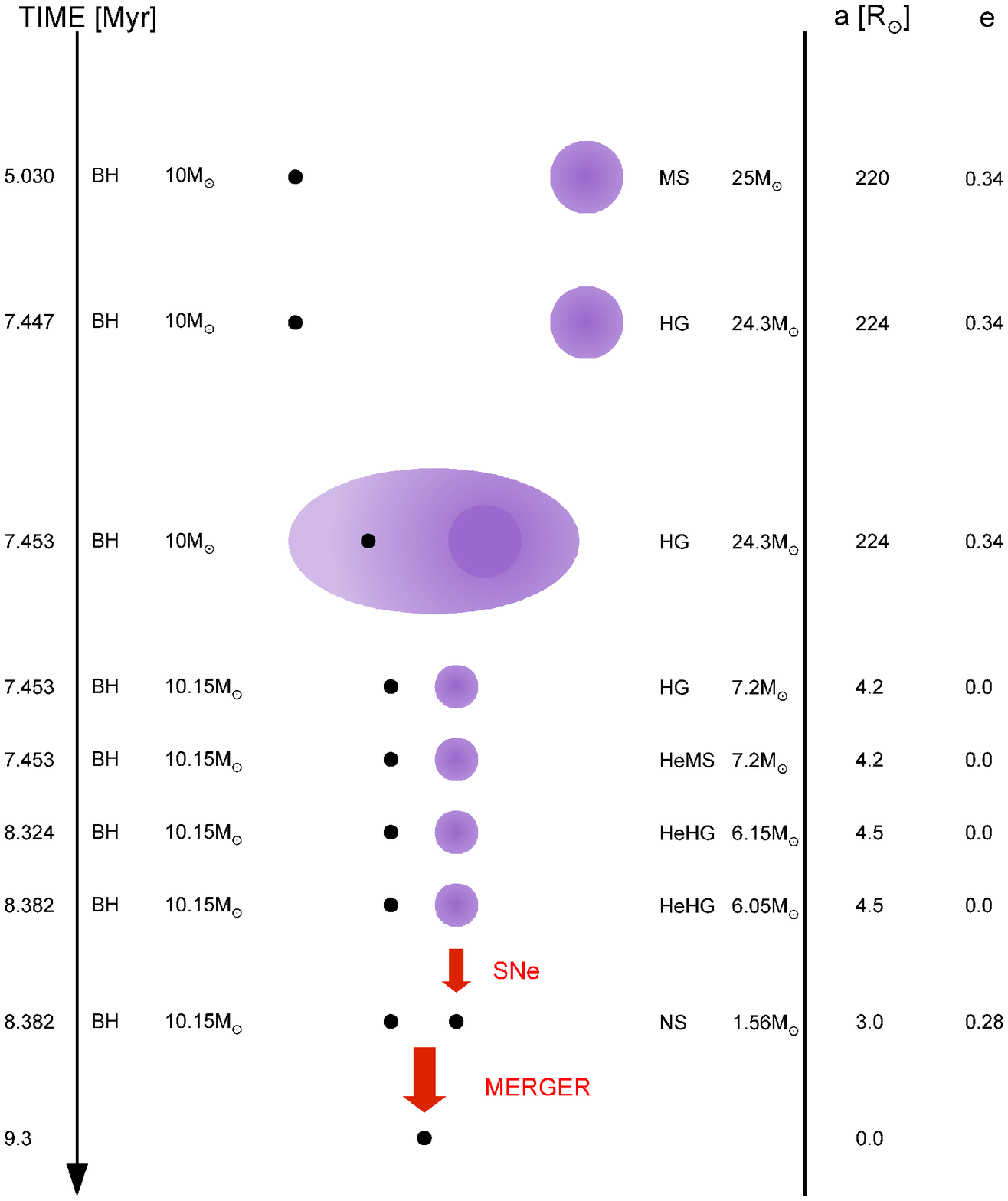}
\FigCap{We start evolutionary calculations at $t=5.03$ Myr, while the more massive 
star has already formed a BH with mass $M_{\rm BH} = 10\MS$ and with assumption that
its companion star has already undergone significant nuclear burning during
its Main Sequence and its mass is $M_{\rm comp} = 25\MS$. The following evolution 
involves common envelope phase and supernova explosion. In the end a close BH-NS 
binary is formed. The details of this evolutionary sequence are given in Sec.2 (case 1).}
\end{figure}

\newpage

\begin{figure}[htb]
\includegraphics[width=\textwidth,height=\textwidth]{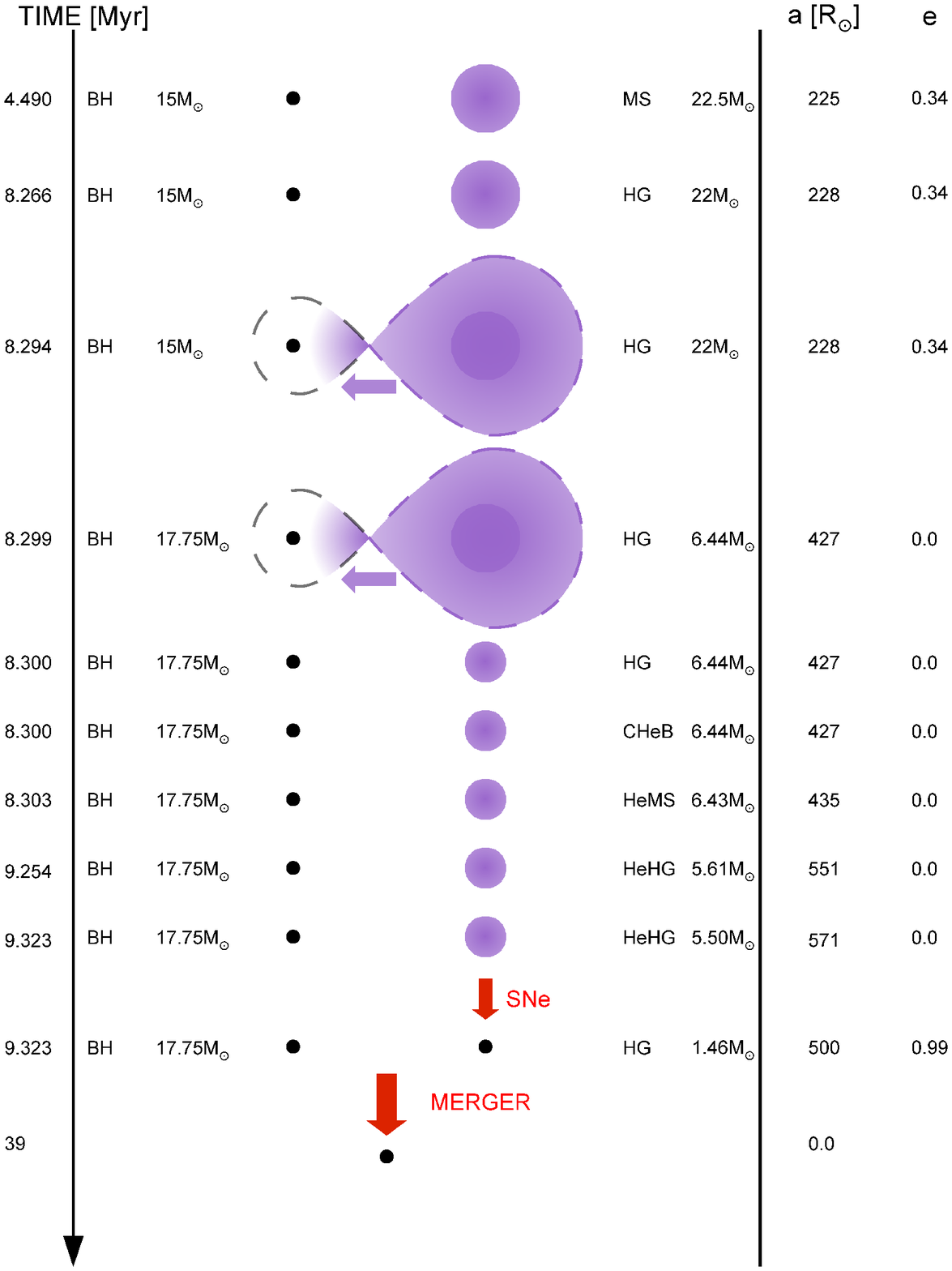}
\FigCap{Typical evolution of P13 along evolutionary channel described in
text as case 2 (see Sec.2 for details).} 
\end{figure}

\newpage

\begin{figure}[htb]
\includegraphics[width=\textwidth,height=\textwidth]{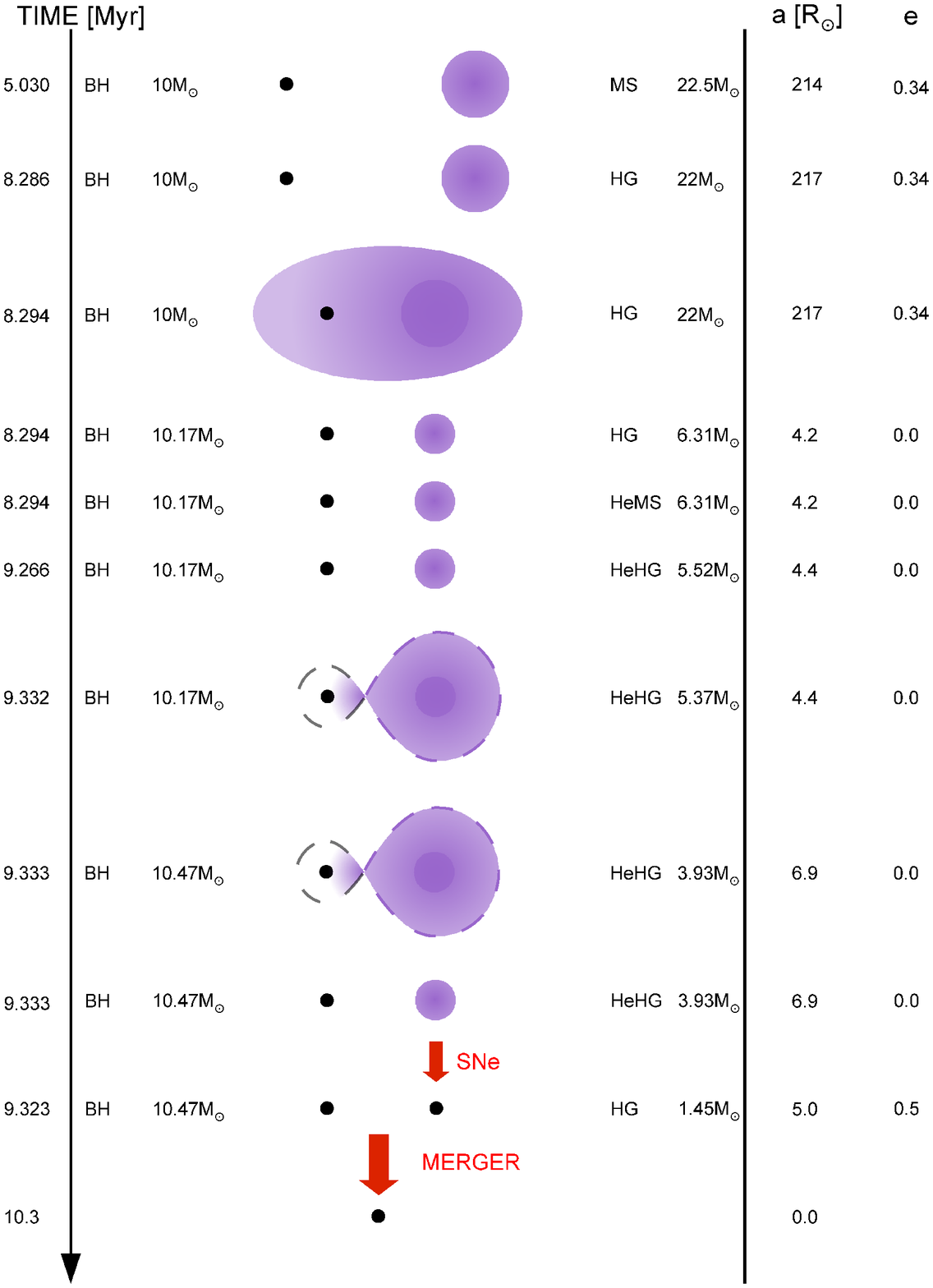}
\FigCap{Typical evolution of P13 along evolutionary channel described in
text as case 3 (see Sec.2 for details).} 
\end{figure}

\end{document}